# Bandwidth-Unlimited, Dispersion-Free Tunable Optical Delay Using a Prism Pair

**Chil-Min Kim[1] and Byoung S. Ham[2,\*]**

[1]*CAcceleration Research Center for Quantum Chaos Application, Department of Physics, Sogang University, 1 Sinsu-dong, Mapo-gu, Seoul 121-742, S. Korea*
[2]*Center for Photon Information Processing, and the Graduate School of Information and Telecommunications Inha University, 253 Yonghyun-dong, Nam-gu, Incheon 402-751, S. Korea.*
\* bham@inha.ac.kr

**Abstract:** We report a nearly perfect bandwidth-unlimited, dispersion-free tunable optical delay system using a prism pair. The observed delay-bandwidth product using a 25-femtosecond light pulse is $\sim 7 \times 10^4$, where the delay time is independent of the signal bandwidth. The present technique breaks from the conventional methods constrained by delay-bandwidth product, nontunability, and delay time dependent optical loss.

**Introduction**

Optical buffer memory [1] has been intensively studied for the last decade since the observation of ultraslow light [2] in Bose Einstein condensate via electromagnetically induced transparency (EIT) [3]. Ultraslow light, however limits the delay-bandwidth product, a practical constraint in information processing due to the Kramers Kronig relation given by absorption and dispersion correlation [4-10]. This is because the signal pulse bandwidth determines the maximum delay time. So far the observed delay-bandwidth product using the ultraslow light is nearly unity [2,4,10]. In a stopped light regime [5,11] or in a microresonator chain scheme, this delay-bandwidth constraint can be lifted. However, optical loss due to dephasing [11] or absorption [1] is inevitable. Thus, the constraints of the delay-bandwidth product and delay-time dependent optical loss must be resolved to implement a tunable optical delay system, a key element of future all-optical signal processing. In this Article we report observations of a nearly bandwidth unlimited, dispersion-free tunable optical delay with no loss using a prism pair. The observed delay-bandwidth product resulting from this method using a 25 fs infrared light pulse is $\sim 7 \times 10^4$ owing to the extended optical path length of 43.4 cm (in vacuum) via controlled multiple total internal reflections (TIR) inside the prism pair. Using a visible light pulse of 115 ns duration (full width at half maximum), we demonstrate an 11 ns optical delay, where the delay time is controllable by simply adjusting the notch length of the prism pair. This prism-pair-based optical delay method is tunable in both frequency and time and nearly dispersionless, with no loss in optical power. Thus, the scheme can be applied to active optical buffer memories in picosecond fiber-optic communication networks.

TIR is an optical phenomenon, which occurs on the interface between two adjacent refractive index materials. In principle no energy loss occurs during TIR. A simple example of TIR involves a circular or spherical microcavity laser which emits light of extremely high quality factor resonance mode, named whispering gallery mode [12,13]. TIR confines whispering gallery mode in the cavity. Based on the whispering gallery mode, slightly deformed as well as completely chaotic shaped cavities have been studied extensively in the context of ray dynamics to improve the directionality of the lasing modes [14,15]. Following this trend in microcavity lasers, our pseudo-integrable optical delay is designed to achieve a maximal traversing length between the right notch and the left notch of the prism pair cavity, i.e., initially entering from the right notch of the prism pair and traveling in a clockwise direction via TIR, but leaving from the left notch in a counterclockwise direction after sweeping the whole cavity area of the prism pair, while the incident angles of the rays between reflections remain intact for TIR.

## 2. Theory

The present tunable optical delay design is similar to the mcrocavity laser with respect to the use of TIR. In the optical design, even thought the light trajectory by TIR is refractive invariant, the total path length is refractive index dependent. The optical path length by TIR inside the optical medium, however, is nearly the same within a few hundred nanometer range from ultraviolet to near infrared. Thus, a prism pair can be used for nearly dispersion-free optical delay based on multiple TIRs. The key advantage of using the prism pair for optical delay is the absence of energy loss or dispersion. Further, the delay time in this prism pair is controllable by adjusting the notch length of the prism pair (discussed in Fig. 1). Additionally, the delay time by the prism pair can be extended by adding another prism pair in series as in the microcavity ring resonator chain [16]. Finally, the present prism pair-based optical delay method can be applied for microphotonics using semiconductors or polymers. Thus, the present method of prism pair-based optical delay has potential for active optical buffer memories, where the delay time can be controllable and an achievable delay-bandwidth product can be a few hundreds or more.

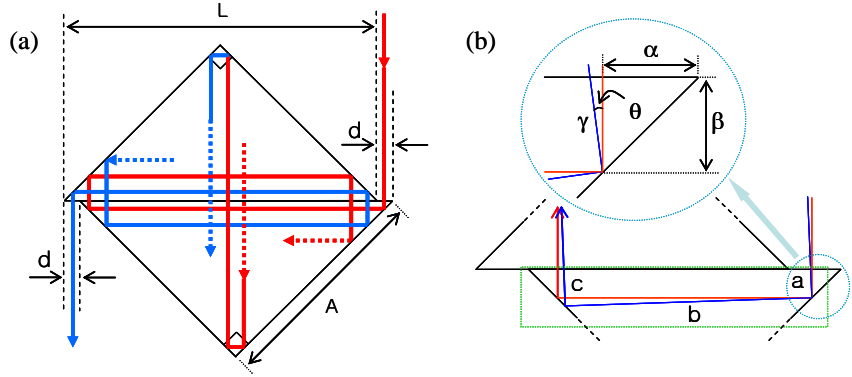

Fig. 1. Schematic of optical delay system using a prism pair for (a) vertical incidence and for (b) path difference diagram between vertical and angled incidence.

Figure 1 shows the schematic of the present prism pair optical delay. Two right angled prisms touched along the inclined plain to form a tilted prism pair. The mismatch, "d" or notch out of the inclined length "L" is an active parameter to control the optical path length "$\ell$":

$$\ell = nNL/2 = \frac{nL^2}{d}, \tag{1}$$

where $n$ is the refractive index, and $N = \frac{2L}{d}$. Since the number N stands for the number of TIRs, which is twice the number of crossing over the interface (see Fig. 1(a)), N should be an even number. From a geometric analysis of the prism pair, each crossover happens with consecutive double TIRs in each side of the prism. This crossover between the prism pair is the only source of optical loss. However the reflection-caused optical loss can be eliminated with an antireflection coating. Here, any two consecutive TIRs in a prism make the same path length $nL$. Thus, the optical delay time $\tau$ by the prism pair is

$$\tau = \frac{n}{c}\frac{L^2}{d}, \qquad (2)$$

where c is the speed of light in vacuum.

In the present prism pair optical delay technique, the delay-bandwidth product depends on the beam path difference caused by the beam divergence. For the beam path $\ell$ and the divergence angle $\theta$, the beam path difference of the ray for $\theta$ against the zero divergence ray (vertical incidence) is $\sim n\ell\theta^2/2$. Then the delay difference between them is $\Delta\tau = \frac{n\ell\theta^2}{2c}$. This path difference is illustrated in Fig. 1b. In normal incidence, $\ell = a+b+c$, unlike angled incidence. For each segment (e.g., segment '$a$' as shown in the inset of Fig. 1(b)), the path length difference is $n\beta(1/\cos(\theta)-1) \sim n\beta\theta^2/2$, where $\theta \ll 1$ and $\beta=\alpha$. This relation applies to all segments. Thus the path length difference per double TIRs or per each crossover, is $\sim nL\theta^2/2$.

Figure 2 shows a numerical simulation of the output delay for a delta input pulse having a spatial Gaussian beam shape with 1 milliradian divergence angle along the notch d. To estimate the distribution of the optical delay time for the traversing lights in the prism pair, we treated the light in Gaussian intensity distributed whose initial maximum divergence angle is 1 miliradian at the right notch of the prism pair, and the beam size is s=0.2d (details are discussed in Fig. 3). The input pulse has, however, no temporal broadening (a delta function) when it enters into the right notch. The light is vertically entering into the middle of the right notch of the prism pair and exits through the left notch as shown in Fig. 1(a). For the simulation we decompose the light of the Gaussian profile by using randomly distributed 1,000,000 rays up to the given maximum divergence angle of 1 miliradian. The path lengths of these 1,000,000 rays are individually calculated for L/d = 20. Then the delayed lengths were plotted for the normalized length L=2. The output path length is obtained by using a histogram. We assumed n=1. In the parameters we set L=2, d=0.1, and the beam size s= 0.2d. Since a commercial laser has less than 1 milliradian in divergence, we set the maximum divergence angle for the beam at the same value of 1 miliradian. The input beam is Gaussian distributed across s.

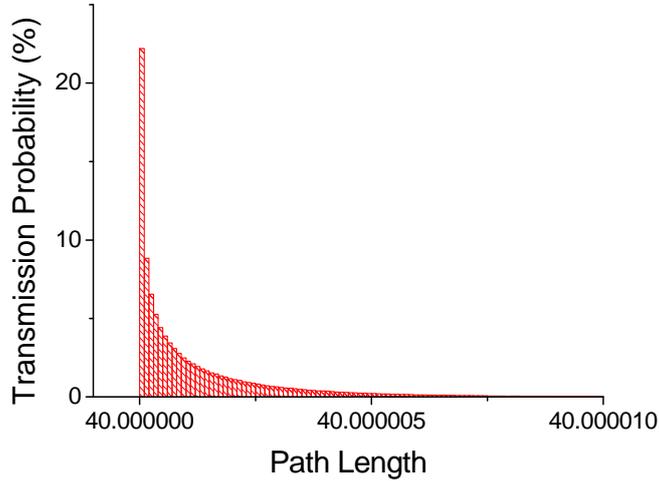

Fig.2. Numerical simulations of Fig. 1. For Eq. (1), path length is $\ell$ =40: L=2, d=0.1, and n=1. The path length is obtained by using 1,000,000 rays for L/d = 20. The input pulse forms a Gaussian profile in the middle of the notch d. The beam size and the maximum divergence are 0.2d and 1 miliradian, respectively.

Figure 3 shows transmission of randomly generated Gaussian profile composed of one million rays as functions of beam size and beam divergence. Under these parameters the dispersion due to the beam divergence is practically negligible in the present optical delay method. As shown in Fig. 2, the temporal delay of the input pulse occurs due to the beam divergence as explained in Fig. 1(b), but the value is less then one millionth. Thus, the effect of a small divergence of the signal light is negligible.

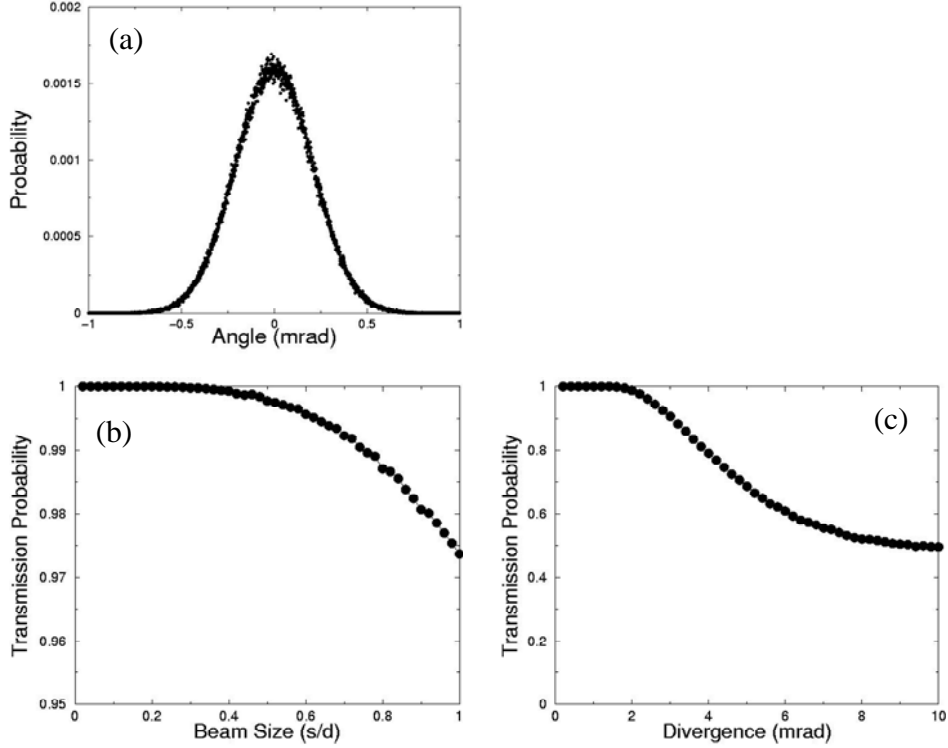

Fig. 3. (a) Randomly distributed Gaussian profile for the input beam. 1 million rays are chosen. (b) Transmission of Gaussian distributed light of (a) as a function of beam size. In the simulation, the beam size is increased up to the notch length for 1 miliradian maximum. The s/d is the ratio of the beam size to the notch length. Even for the full size (s/d=1), the transmission efficiency is higher than 97%. (c) Transmission of Gaussian distributed light of (a) as a function of beam divergence. Up to 2 milliraidans, the transmission is nearly 100%. Even for 10 milliradians, the transmission reduction is just 50%, which is beam divergence tolerable.

## 3. Experimental results and discussions

Figure 4 shows an experimental observation of Fig. 1 for normal incidence, where the optical pulse duration is 115 ns (full width at half maximum), and the wavelength is 606 nm. The signal light is from a ring dye laser at $\lambda=606$ nm (Tecknoscan) whose optical bandwidth is ~300 kHz. We used an acousto-optic modulator (ISOMET) driven by a digital delay generator (DDG 535) and a microwave synthesizer (PTS 250) to make a pulse of 115 ns. The repetition rate is 50 Hz, and the output signal is measured by an avalanche photo diode. For Fig. 4, the light is from a Ti:S femtosecond laser (Chinook, KM Laboratory at 80 MHz repetition rate) at $\lambda=800$ nm. The optical pulse spectrum of the inset of Fig. 4 is measured by using an optical spectrometer (SM 240). A Michelson interferometer was adapted for the second harmonic generation. We roughly set the notch size d in Fig. 4 at a predetermined value according to Eq. (2). The dimension of the anti-reflection coated prism is A=2" (L=7.18 cm). The data in Fig. 4 is directly obtained from a digital oscilloscope by averaging 30 signals. The measured optical delay time by the prism pair is 11 ns, where the optical path length is calculated to be 330 cm in vacuum. Because the material of the prism is BK7 (n=1.515 at λ=606 nm), the actual delayed path length inside the prism pair should be 218 cm. From Eq. (1), we calculate that the total number of TIRs is N=60, and the notch is d=2.37 mm. In the experiment, we could not measure d exactly, but chose an arbitrary mismatch d suitable for the laser beam, whose beam diameter is approximately 1 mm. The optical delay time is determined only by the prism pair geometry with d as shown in Eq. (2). This means that the delay time is controllable by adjusting the notch length d; the smaller the notch d, the longer the delay time τ.

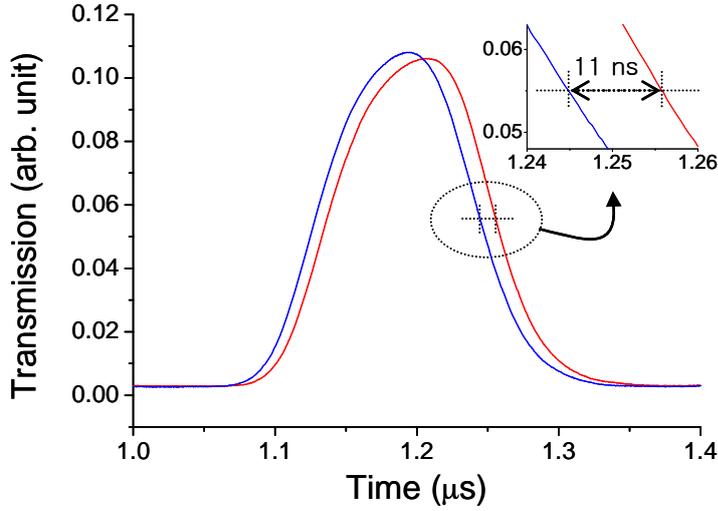

Fig. 4. Observation of optical delay using the prism pair in visible regime. The blue curve is for a reference pulse at 115 ns pulse duration as a reference bypassed the prism pair. The red curve is the delayed pulse by 2" prism pair.

Figure 5 shows an optical delay using a 25 fs optical pulse (λ=800 nm) in the prism pair of Fig. 1. For Fig. 5, we use a 1" anti-reflection coated prism pair (L=3.59 cm; d=4.49 mm) for N=16, where the optical path length in vacuum is measured at 43.4 cm (see Experimental Methods). Because no photo detector can measure a 25 fs light pulse, we use a Beta Barium Borate (BBO) crystal for second harmonic generation. To do this, initially both reference (R) and probe (S) beams are focused onto the BBO crystal to confirm coincidence of the two pulses by detecting second harmonic generation, adjusting both Michelson arms A and B in Fig. 5. Then the prism pair is inserted to the probe line S, roughly decreasing the arm length "B", and fine tuning the reference arm "A" until the second harmonic generation is observed (see the blue spot $\chi^{(2)}$ between S and R on the screen in Fig. 5). The length difference measured is 43.4 cm with a delay time of τ=1.45 ns. The refractive index n of the BBO crystal for λ=800 nm is n=1.510; thus the actual path length delay inside the prism pair should be $\ell/n$=28.7 cm. As shown in the inset of Fig. 4, the measured spectrum (black curve) of the signal light before entering the prism pair is ~50 THz. The measured spectrum (red curve) after the prism pair narrows due to pulse broadening, as expected. From these values, the delay-bandwidth product is DBP = $(50 \times 10^{12})(1.45 \times 10^{-9})$ = $7.25 \times 10^4$.

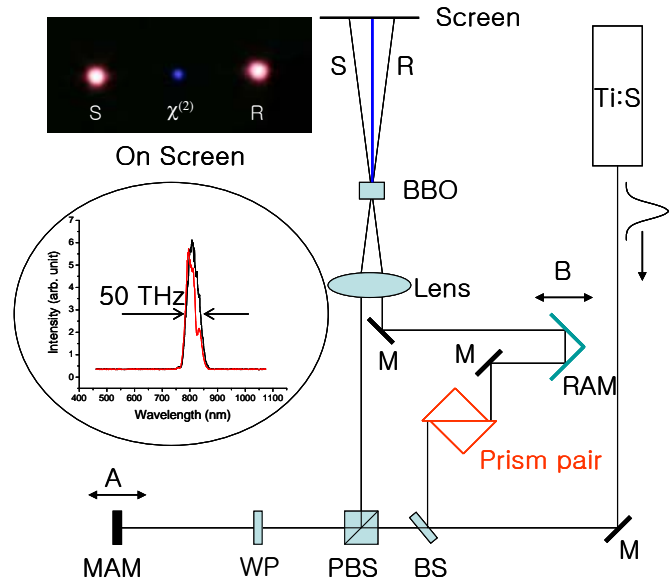

Fig. 5. Observation of optical delay using a 25 fs light pulse in a prism pair. Inset: Optical spectrum of the input and output pulses across the prism pair. M: Mirror, BS: beam splitter, MAM: micromotor activated mirror, RAM: right angled mirror, PBS: polarization beam splitter, WP: λ/4 wave plate.

### 3. Conclusion

In conclusion, we reported a simple but powerful method of a nearly bandwidth unlimited, dispersion-free tunable optical delay system using a prism pair. The observed delay-bandwidth product is $7\times10^4$ and holds potential for tunable optical buffer memories in picosecond fiber-optic communication networks.

**Acknowledgement**

This work was supported by Acceleration Research Program (Center for Optical Chaos Applications) and Creative Research Initiative Program (Center for Photon Information Processing) of MEST via National Research Foundation, S. Korea. We thank S. M. Ma for the experimental measurement of the second harmonic generation in Fig. 5.